\documentclass[11pt]{article}
\usepackage{geometry}                
\usepackage{graphicx}
\usepackage{amssymb,amsmath}
\usepackage{epstopdf}
\usepackage{color}
\newcommand{\G}{\gamma}
\newcommand{\GP}{\gamma^\prime}
\newcommand{\too}{\rightarrow}
\newcommand{\mix}{\chi}
\newcommand{\mixo}{\chi_{_0}}
\newcommand{\OP}{\omega_\mathrm{P}}

\newcommand{\muu}{ \mu } 
\newcommand{\BSP}{\hspace{1cm};\hspace{1cm}}

\newcommand{\MV}{ v_{_\mathrm{M\ddot{o}l}    } }

\renewcommand\({\left(}
\renewcommand\){\right)}
\renewcommand\[{\left[}
\renewcommand\]{\right]}

\newcommand{\dd}{{\rm d}}

\def\be{\begin{equation}}
\def\ee{\end{equation}}
\def\bea{\begin{eqnarray}}
\def\eea{\end{eqnarray}}

\newcommand\mpl{m_{\rm p}}
\newcommand{\mcL}{{\mathcal L}}

\newcommand{\GeV}{{\rm GeV}}
\newcommand{\MeV}{{\rm MeV}}
\newcommand{\keV}{{\rm keV}}
\newcommand{\abs}{\rm D}

\title{
\begin{flushright}
\small NIKHEF/2008-030\\
DESY 08-154
\end{flushright}
\Large{\textbf{Massive Hidden Photons as Lukewarm Dark Matter}} }

\author{
Javier Redondo$^{1}$
and Marieke Postma$^{2}$
\\[2ex]
\small{\em $^1$Deutsches Elektronen-Synchrotron, Notkestra\ss e 85, 22607 Hamburg, Germany}\\
\small{\em $^2$Nikhef, Kruislaan 409, 1098 SJ Amsterdam, The Netherlands.}
}
\date{}

\begin{document}

\maketitle

\begin{abstract}
\noindent
We study the possibility that a keV-MeV mass hidden photon (HP), i.e. a hidden sector U(1) gauge boson, accounts for the observed amount of dark matter. We focus on the case where the HP interacts with the standard model sector only through kinetic mixing with the photon.
The relic abundance is computed including all relevant plasma effects into the photon's self-energy, which leads to a resonant yield almost independent of the HP mass. The HP can decay into three photons. Moreover, if light enough it can be copiously produced in stars. Including bounds from cosmic photon backgrounds and stellar evolution, we find that the hidden photon can only give a subdominant contribution to the dark matter. This negative conclusion may be avoided if another production mechanism besides kinetic mixing is operative.
\end{abstract}


\section{Introduction}

The evidence for the existence of dark matter (DM) is compelling\cite{Jungman:1995df,Bertone:2004pz}.  The various measurements, using rotation curves of galaxies, lensing data, cluster dynamics, large scale structure and CMB data all agree: about 25\% of the energy budget of the Universe is in the form of dark matter, and a large part of the dark matter is non-baryonic. Although observational evidence for dark matter is plentiful, we are still in the dark about its identity. The theory of structure formation provides indirect evidence about some of its properties. It favors cold dark matter (CDM) that is weakly interacting and non-relativistic at late times.   One of the best motivated CDM candidate is a WIMP, a weakly interacting massive particle \cite{Lee:1977ua,Vysotsky:1977pe}. The ligthest supersymmetric particle is the archetypical WIMP example. WIMPs are in thermal equilibrium with the standard model (SM) particles in the early universe. With weak scale masses and interactions WIMPs would have fallen out of equilibrium at the right time such that their relic density today is in the right 
ballpark.

Although CDM provides a consistent picture of structure formation on large scales, there are persistent problems on subgalactic scales. Most notably, CDM predicts too many galactic satellites \cite{Kauffmann:1993gv,Klypin:1999uc,Moore:1999nt,Diemand:2006ik}, a galaxy density profile that is too cuspy \cite{Moore:1994yx,deBlok:2001fe,Gentile:2004tb,Simon:2004sr}, and too low angular momenta of spiral galaxies \cite{SommerLarsen:1999jx}. All these problems suggest that CDM may be too cold. This has motivated warm dark matter (WDM) \cite{Bode:2000gq,AvilaReese:2000hg}.  WDM consists of lighter particles, in the 1-10 keV range, that are on the borderline between non-relativistic and relativistic at the time of structure formation. The non-zero velocity dispersion suppresses structure below the Mpc scale. As a result structure on sub-galactic scales is damped, resolving the tension between DM matter simulations and observations. Note in this respect that even lighter particles would  still be relativistic at late times; such hot dark matter damps structure on scales much larger than galaxy scales, in conflict with observations.

The standard model of particle physics (SM), while describing collider phenomenology very successfully, does not offer a suitable DM candidate, and we must look further. Most SM extensions include hidden sectors, i.e. sectors which couple only very weakly, typically gravitationally, to the SM fields. Given the importance of symmetries in the SM it is not unlikely that these hidden sectors contain gauge groups as well, including some abelian U(1) factors. The usual assumption is that such hidden photons (HP) are heavy, and decouple from the standard model particles, thereby avoiding all observational constraints. But this needs not be the case. For example, if the hidden U(1) is broken by non-perturbative effects, the symmetry breaking scale and thus the photon mass is exponentially suppressed, and can be naturally light \cite{Burgess:2008ri}. 
At low energies the dominant coupling of the hidden photon to the SM will most likely be via kinetic mixing with the SM photon.  Kinetic mixing is allowed by all symmetries, and can be realized via a renormalizable coupling --- thus in principle unsuppressed by heavy mass scales --- of the form $\mcL \ni \chi F_{\mu \nu} B^{\mu \nu}$, where $F_{\mu \nu}$ and $B_{\mu \nu}$ are the SM and hidden photon field strengths respectively~\cite{Holdom:1985ag}. 

In this paper we study the possibility that a hidden photon with mass in the keV-MeV range, and whose dominant coupling to the standard model is via kinetic mixing, is a dark matter candidate. Because ``warm" is normally used for thermal relics with masses in the $1-10$ keV, our keV-MeV candidates are more likely ``lukewarm". 
Such hidden photons are too heavy to be measured in laboratory experiments; 
the 5th force searches or the laser experiments designed to search for axion like particles are only sensitive to much lighter hidden photons~\cite{Bartlett:1988yy,Ahlers:2007qf,Ahlers:2007rd}. Nevertheless, there are astrophysical and cosmological constraints that restrict this scenario considerably. First, there is the requirement that hidden photons (HP) do not overclose the universe. In calculating the HP relic density we focus on the case where HPs are solely produced through their mixing with photons, although we comment on other possibilities. Second, such light and weakly coupling particles are produced in the core of stars, and can subsequently escape the star unimpeded, providing an extremely efficient cooling mechanism. This alters the evolution of stars, which is bounded by observations. These constraints are strong. Indeed, in the whole mass range below 100 keV the mixing parameter for which interesting (measurable) abundances are obtained is ruled out.
Third, the HP is an example of decaying dark matter, as it can decay into three visible photons. If it is to be the dark matter its lifetime should be larger than the age of the universe, which bounds the kinetic mixing from above. And fourth, the decay products of the HP contribute to the galactic and cosmic gamma-ray background, again constrained by observations. As we will see, all constraints put together practically rule out our hidden photon dark matter scenario, unless another production mechanism besides kinetic mixing is operative.

Under the assumption that kinetic mixing is the dominant production mechanism of hidden photons in the early universe, the hidden photon cannot be a cold dark matter candidate with MeV mass or above. The reason is that for such heavy HPs the decay channel into an electron-positron pair is open. Stability on the scale of the lifetime of the universe can only be obtained for an extremely small kinetic mixing. However then the relic abundance is way too small to be dark matter. 
This raises the question whether such super-MeV mass HP can leave any imprint on cosmology at all. 
If the HPs decay after the time of nucleosynthesis and/or matter radiation decoupling (CMB), can they leave measurable traces? 
Also here we find a negative answer. 

While this work was in progress two related papers appeared. Pospelov et al. \cite{Pospelov:2008jk} also discuss the possibility of hidden photons as dark matter. Although there is a large overlap, we extend their analysis by including resonance effects (which actually dominate production), and by significantly improving the stellar bounds. As a result, our conclusions differ substantially from~\cite{Pospelov:2008jk}. In particular, whereas they find some hidden photon mass and kinetic mixing parameters for which the HP can be the dominant source of dark matter, we rule out this possibility. Chen et al. \cite{Chen:2008yi} study HPs with masses of $\sim 100$ GeV as cold dark matter. Their analysis differs from ours in that they include additional non-renormalizable interactions, which provide an extra source of production. This allows for a long lifetime (requiring small kinetic mixing) while at the same time obtaining a large relic density.

This paper is organized as follows. In the next section we introduce the model. For a proper treatment of photon - hidden photon interactions plasma effects should be taken into account and the relevant formulas are discussed. In section~\ref{s:production} we calculate the relic density of hidden photons as a function of its mass and mixing.  We explain that for hidden photons in the keV range production is dominated by the resonance regime, which occurs for temperatures such that the thermally induced photon mass equals the hidden photon mass.
For heavier HPs the main production channel is via electron-positron coalescence. In section~\ref{s:bounds} we discuss the cosmological and astrophysical bounds on the HP parameter space, including the bounds from overclosure, stellar evolution, and from the diffuse gamma-ray background. In section \ref{s:non-ren} we discuss the effects of non-renormalizable interactions between the hidden sector and the standard model. If these additional interactions dominate the production of hidden photons parameter space opens up, and hidden photons can be the dark matter in the universe. We end with some concluding remarks.


\section{The model}\label{s:model}

Consider a hidden sector U(1) gauge boson, a hidden photon, which couples to the standard model photon solely via gauge kinetic mixing~\cite{Holdom:1985ag} with the hypercharge boson. 
The low-energy effective Lagrangian is~\cite{Redondo:2008aa,Jaeckel:2008fi}
\be
\mathcal{L} =
-\frac{1}{4}F_{\mu \nu} F^{\mu \nu}
 - \frac{1}{4}B_{\mu \nu} B^{\mu \nu}
+ \frac{\sin\mixo}{2} B_{\mu \nu} F^{\mu \nu}
+ \frac{\cos^2\mixo}{2} \muu^2 B_{\mu} B^{\mu},
\label{LagKM}
\ee
where $F_{\mu\nu}$ and $B_{\mu\nu}$ are the photon ($A^\nu$) and
hidden photon ($B^\nu$) field strengths.  The dimensionless mixing
parameter $\sin\mixo$ can be generated at an arbitrarily high energy
scale and does not suffer from any kind of mass suppression from the
messenger particles communicating between the visible and the hidden
sector. This makes it an extremely powerful probe of high scale
physics. Typical predicted values for $\mixo$ in realistic string
compactifications range between $10^{-16}$ and
$10^{-2}$~\cite{Abel:2003ue,Abel:2006qt,Abel:2008ai,Dienes:1996zr}.

The most prominent implication of the kinetic mixing term together with the non-zero hidden photon mass $\mu$ is that
photons are no longer massless propagation modes.  Similar to neutrino
mixing, the propagation and the interaction eigenstates are misaligned.
The kinetic mixing term can be removed by a change of basis
$\{A,B\}\rightarrow\{A_{_R},S\}$, where $A_{_R}=\cos\mixo A$ and
$S=B-\sin\mixo A$.  Since $A$ and $A_{_R}$ differ only by an
unobservable charge renormalization we will drop the $R$ subscript
from now on. In the $\{A,S\}$ basis the kinetic term is diagonal but
kinetic mixing provides an off-diagonal term in the mass-squared
matrix,
\begin{equation}
\left(
\begin{array}{cc}
\muu^2 \sin\mixo^2 & \muu^2  \sin\mixo\cos\mixo     \\
\muu^2\sin\mixo\cos\mixo   & \muu^2 \cos^2\mixo
\end{array}
\right) \ \ .
\end{equation}
As a result one expects vacuum photon-sterile
oscillations~\cite{Okun:1982xi} as in the case of neutrinos.  In
the following we will use the notation $\G$ ,$\GP$ for the
flavor states, which are the quanta of the $\{A,S\}$-fields respectively.  The mass
eigenstates are denoted by $\gamma_{1,2}$, with $\gamma_1$ mostly
``photon-like'' and $\gamma_2$ mostly ``hidden photon-like''.

The above discussion applies to the system in vacuum. In the early
universe matter effects should be taken into account, as the photon is
submerged in a thermal plasma. We can include the influence
of the plasma in the photon propagation\footnote{
Throughout this paper we do not discuss longitudinal photons.  We expect that including their effects will not affect our results significantly.}
through the photon's self energy, which enters into the above formalism as a complex effective mass. The real part of the mass encodes the refraction properties of the plasma and its magnitude is set by the plasma frequency $\OP$. We will refer to it as the plasma mass or the photon mass denoted by $m_\gamma$. For the temperatures of interest the main contribution comes from Compton scattering on electrons, which gives
\be
m_\gamma^2 = 
\left\{
\begin{array}{ll}
\omega_p^2 = 4\pi \alpha (n_e/m_e),
& \qquad(T \ll m_e) \\
\frac32 \omega_p^2 = (2/3)\alpha \pi T^2 ,
& \qquad (T \gg m_e) 
\end{array}
\right.
\label{mgamma}
\ee
where $n_e$ is the electron number density. The contributions from photon scattering off other charged particles are completely analogous. The full formulas are given in Appendix~\ref{A:wp}.

The imaginary part of the photon effective mass is given by
\be
\omega \abs(\omega,T) =-\omega\(\Gamma^{\rm A}(\omega,T)-\Gamma^{\rm P}(\omega,T)\)=-\omega\(e^{\omega/T}-1\)\Gamma^{\rm P}(\omega,T)
\label{abs}
\ee
where $\Gamma^{\rm P(A)}(\omega,T)$ is the production(absorption) rate of photons with energy $\omega$ in a thermal bath at temperature $T$~\cite{Weldon:1983jn,Redondo:2008aa}. The ``damping factor'' $\abs(\omega,T)$ may be thought of as a rate parameter measuring the effectiveness of the collisions to stop the coherent development of the wave function \cite{Stodolsky:1986dx}; alternatively it can be interpreted as the rate at which photons would regain thermal equilibrium~\cite{Weldon:1983jn}. The relevant definitions are presented in Appendix~\ref{A:wp}.

The effective mixing angle in a damping dominated medium\footnote{For the small values of $\mixo$ we consider in this paper, the mixing is always
damping dominated with $\omega \abs \gg 2 \mixo^2 \muu^2$} is given by \cite{Redondo:2008aa} 
\begin{equation}
\mix^2(\omega,T) \simeq \mixo^2 \frac{\muu^4}{(\muu^2-m_\gamma^2)^2+(\omega \abs)^2} \ . 
\label{mix}
\end{equation}
It depends implicitly on the energy and the temperature though $m_\G$ and $\abs$.
The imaginary contribution to the photon mass, $\omega\abs$, is typically smaller than the real part\footnote{Both the real and imaginary parts come from matrix elements squared, but the real part interferes with the identity matrix and therefore involves less powers of the coupling constant.}, $m_\gamma$, so it only plays a role near the resonance $m_\gamma=\muu$ where it acts as a cut off.

Since the plasma frequency is a steep function of temperature, the resonant condition $m_\gamma(T_r)=\muu$ divides the temperature range into a region  of very suppressed mixing angles ($m_\gamma\gg \muu$ for $T\gg T_r$) and a region which can be considered as vacuum ($m_\gamma\ll \muu$ for $T\ll T_r $). For heavy hidden photons with $\mu \gtrsim \MeV$, the resonance happens when electrons are still relativistic at a temperature $T_r= \muu\sqrt{3/(2\pi \alpha)}\simeq 8 \muu$, where we used \eqref{mgamma}. For much smaller masses the  electrons are non-relativistic at the time of resonance, and $m_\gamma \propto T^{3/2} e^{-m_e/T}$. Due to the exponential decrease the thermal mass $m_\gamma$ is extremely sensitive to $T$. Consequently for a broad range of values for $\muu$, in the range $1-10^5$ eV, the resonance happens not far from $ T_r \sim 0.2\ m_e$.  
As we will see, for small HP masses the production is dominated by the resonance, whereas for larger masses at slightly lower temperatures. This means that in both cases all interesting physics happens at relatively high temperatures $T > 0.2\ m_e$, corresponding to the period when electrons and positrons have not completely annihilated and their chemical potential is still negligible.

That the resonance is more prominent for low mass HP masses can be easily understood. Considering only Compton scattering the damping factor $\abs$ ranges from $8 \pi \alpha^2/(3m_e^2)n_e$ (at $T\ll m_e$) \cite{Redondo:2008aa} to $\sim \alpha^2 T^2/(\pi\omega)$Log$(4 T \omega/m_e^2)$ (at $T\gg m_e$). This gives for the ratio
\be
\frac{m_\gamma^2}{\omega \abs}\simeq 
\left\{
\begin{array}{ll}
 \frac{3}{2\alpha}m_e/\omega,
& \qquad(T \ll m_e) \\
& \\
 \frac{2 \pi^2}{3 \alpha}\({\rm Log}\frac{4 T \omega}{m_e^2}\)^{-1},
& \qquad (T \gg m_e) 
\end{array}
\right. 
\label{ratio}
\ee
which sets the enhancement of the effective mixing angle at the resonance. Since $\omega\sim T$, \eqref{ratio} shows clearly how the resonance is enhanced at low temperatures, corresponding to small HP masses, and losses importance (first linearly, thereafter logarithmically) as electrons become more and more relativistic. For our range of hidden photon masses $\muu >$ keV, the resonance temperature is bounded $T_r \gtrsim 0.2\ m_e$, and the enhancement factor \eqref{ratio} is never extremely large, a factor $\sim 10^3$ at most.  

The photon-hidden photon oscillation frequency also becomes modified in the medium, it reads 
\begin{equation}
\omega_{\rm osc}=\frac{1}{2 \omega}\sqrt{(\muu^2-m_\gamma^2)^2+4\mixo^2\muu^2}  \ \ \ . 
\end{equation}
Since typically $\omega_{\rm osc}\gg \abs$, many oscillations take place before a photon is absorbed or scattered so the two mass eigenstates, which travel at different speeds, are not likely to still have an overlap at subsequent interactions. We can then simply treat $\gamma_{1,2}$ as two different final states. This is true except near the resonance, where $\omega_{\rm osc}/ \abs$ is precisely the ratio in \eqref{ratio} times the vacuum mixing angle $\mixo$, and therefore is very small unless $\mixo\gtrsim 2\alpha \omega/(3 m_e)$. 
However, even in this last case we can also treat $\gamma_{1,2}$ as two final states (neglecting coherence) \emph{as long as the effective mixing is small} because the photon-like wave/state/component $\gamma_1$ will be damped much faster than $\gamma_2$.


\section{Production of hidden photons}\label{s:production}

In this section we calculate the abundances of thermally produced
hidden photons.  We can divide the production into three stages.  At high temperatures
for which $m_\gamma(T) \gg \muu$ the hidden photon is comparatively
massless; photons are very close to being both interaction and
propagation eigenstates and the effective mixing angle \eqref{mix} is strongly 
suppressed.  Consequently the amount of hidden photons produced is
negligible small. 
We can therefore assume that the initial abundance of hidden
photons is negligible small. This sets the initial condition. 
 As the temperature lowers, so does the plasma frequency, and the system
hits the resonance when $m_\gamma(T) = \muu$.  The photon bath present
will partly convert into a hidden photon bath, the efficiency of this
conversion depending on the effective mixing parameter. 
 At lower temperatures, the effective mixing relaxes to the vacuum value $\mix\sim \mixo$, but still  the production can be effective, favored by the fact that the expansion of the universe is increasingly slower.  

Since we have argued that coherent effects do not play a role, the evolution equation for the $\gamma_2$ yield is the usual Boltzmann equation one expects from incoherent production,
\be
\frac{\partial Y_2}{\partial \ln T} = 
\frac{\Gamma_2}{H} \times \frac{\dd \ln s}{\dd \ln T^3} Y_1
\label{beq}
\ee
where we defined $Y_{1,2} = n_{1,2}/s$ the ratio of the
hidden photon-like or SM photon-like density to the entropy density, 
and $H$ is the Hubble constant.
The $\dd \ln s$ term on the right hand side incorporates the change in the effective plasma degrees of freedom as species decouple from the thermal bath \cite{Gondolo:1990dk}.
For the broad HP mass range keV-GeV, there is only
muon, electron and neutrino decoupling. The relevant definitions can be found in 
Appendix~\ref{A:dof}.
Note finally that we assumed that $Y_2 \ll 1$ always, making the absorption
processes of $\gamma_2$s negligible.  

The HP production rate is given by the sum of different contributions
\be
\Gamma_{2}= (n_{e^+} +n_{e^-}) \langle \sigma_{\gamma_2 e} \MV \rangle + 
\frac{n_{e^+} n_{e^-}}{n_\gamma} \langle \sigma_{\gamma_2} \MV \rangle +
\frac{n_{e^+} n_{e^-}}{n_\gamma} \langle \sigma_{\gamma_1\gamma_2} \MV \rangle + ...
\ee
where $n_{e^+},n_{e^-}$ are the number densities of electrons and positrons, $\sigma_{\gamma_2 e}, \sigma_{\gamma_2}, \sigma_{\gamma_2\gamma_1}$ are the spin-averaged cross sections of the reactions $\gamma_1 e  \to \gamma_2 e $ (Compton-like production), $e^+ e^- \to \gamma_2$ (pair coalescence) and $e^+ e^- \to \gamma_1 \gamma_2$ (pair annihilation). The brackets denote the proper thermal average, which for a reaction $a+b\to \gamma_2 + c$ is given by~\cite{Gondolo:1990dk}
\footnote{
We neglect stimulation/blocking factors for final state bosons/fermions since in practice they do not play a significant role.} 
\bea
n_a n_b \langle \sigma_{\gamma_2 c} \MV \rangle
&=& \int 
\frac{g_a\dd p_a^3}{(2\pi)^3} f_a
\int \frac{g_b\dd p_b^3}{(2\pi)^3} f_b
\, \sigma_{\gamma_2 c}(s) \, \MV \equiv \int \dd n_a \dd n_b \, \sigma_{\gamma_2 c}(s) \, \MV
\label{thermal1}
\eea
with $f_{a,b}$ the Fermi/Bose distribution function $= ( e^{E_{a,b}/T} \pm 1 )^{-1}$ for initial state fermion/boson, $\MV=\sqrt{|v_a - v_b|^2 - |v_a \times v_b|^2}$, the Moeller velocity,  
$E_{a,b},p_{a,b},v_{a,b}$ the energy, three momenta and three velocity of the incoming particles, $s$ the center-of-mass energy 
(not to be confused with the entropy density) and $g_x$ the internal degrees of freedom of the particle type $x$. 
The relevant cross sections can be found in Appendix~\ref{A:cs}.
At temperatures $T \ll m_e$ only the scattering process contributes to HP production because of the exponentially decreasing electron/positron density. 
At large temperatures however we also have to include the coalescence $e^+ e^- \to \gamma_2$ and annihilation $e^+ e^- \to \gamma_1 \gamma_2$ processes. The coalescence process is much stronger since it is suppressed by a smaller power of the fine structure constant $\alpha\simeq 1/137$ but it is only possible for HP masses greater than twice of the electron, i.e. for $\muu> 2m_e$. Therefore in principle we should consider all three processes\footnote{The generalization to include other charged particles like muons or pions is straightforward, but for the HP mass range of interests can be neglected.}.

To make the numerics less demanding we can use Maxwell-Boltzmann statistics for the initial particles.  Leaving aside the resonant production, for which we will develop a simple and general formula, this is justified because the production is dominated by the Wien tail of the distribution. The reason is that the production rate is proportional to some inverse power of the temperature times a Boltzmann factor $e^{-E/T}$, and thus production is dominated by temperatures $T < \mu$.
Only the particles in the Wien tail of the distributions will have enough energy to produce a HP, and their distribution is well encoded by Maxwell statistics. Note also that in this regime $\muu\gg m_\G$ and this justifies why we can set $m_\G=0$ in the cross sections of in Appendix~\ref{A:cs}. Under these conditions the thermal average \eqref{thermal1} simplifies enormously
\be
\frac{n_a n_b}{n_\gamma} \langle \sigma v_{_\mathrm{M\ddot{o}l}}\rangle = \frac{g_a g_b}{g_\gamma}\frac{1}{32 \pi^2 \zeta (3) T^2}
\int_{s_0}^{\infty} (s-s_0)ds \sqrt{s}\, \sigma(s) K_1\(\frac{\sqrt{s}}{T}\) \ , 
\ee
where $s_0=(m_a+m_b)^2$ and $K_1$ is the modified Bessel function of the second kind. 

\begin{figure}[t]
\begin{center}
\includegraphics[width=16cm]{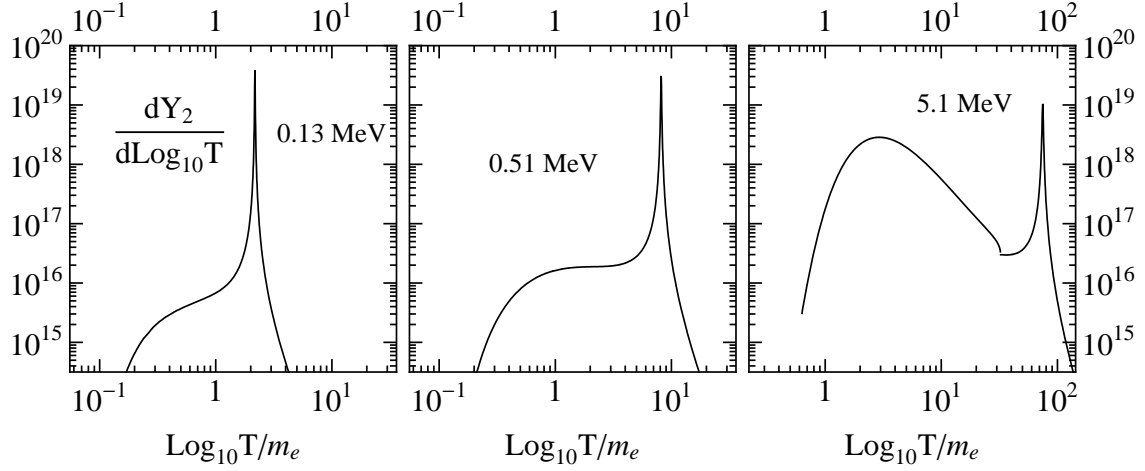}
\caption{Differential production rate of hidden photons for three different masses. For $\muu=10 m_e$ (right) the low temperature production through electron-positron coalescence dominates the final abundance, the resonance not being extremely peaked. Coalescence happens only for $T \lesssim 3\muu$ because for higher temperatures the electron thermal mass is too high. Production of HPs of higher masses proceeds very much in the same way. For $\muu=m_e$ (center) coalescence is never possible, and the low temperature production suffers a ${\cal O}(\alpha)$ decrease since it is now dominated by Compton scattering and pair annihilation. The resonance is more peaked, dominating production. 
Finally for $\muu=m_e/3$ (left) the resonance has grown very large with respect to the low temperature production; the production rate is suppressed by the lack of ambient electrons.}
\label{dY2}
\end{center}
\end{figure}

We have computed numerically this average and plotted our results for three representative cases in Fig.~\ref{dY2}. For illustrative purposes we have extended our calculations to large values of the temperature to cover the resonance as well. Looking at the graph of $\dd Y_2/\dd \log_{10}T$ we see that the resonant contribution grows with respect to the low temperature incoherent part for small HP mass. Indeed, where for $\muu > 2 m_e$ incoherent scattering dominates production, in the opposite limit $\muu < 2 m_e$ resonant production dominates.
For heavy HPs with $\muu > 2 m_e$ the dominant contribution to the final HP abundance
comes from incoherent production via pair coalescence; the resonance contributes only a
small fraction.  We discuss this case in the next subsection.  
Below $\muu < 2 m_e$ the coalescence process is not possible, and the incoherent production decreases substantially compared to the resonance, which is already very peaked and therefore dominates HP production.  More details are given in subsection~\ref{s:resonance}.

\subsection{The coalescence region $(\muu > 2 m_e)$}

For heavy HPs with mass $\muu > 2 m_e$  pair coalescence dominates.
Compton-like and annihilation contributions are subleading because of the additional power of $\alpha$ in the cross sections. Moreover, even though they contribute to the resonance (unlike coalescence) this contribution remains less than $\sim 10$ \%.

The coalescence cross section contains a Dirac delta of the center-of-mass energy which makes the thermal average trivial. We find ($n_{e^+}=n_{e^-}=n_e$)
\be
\frac{n_e^2}{n_\gamma}\langle \sigma_{\gamma_2} \MV\rangle = 
\frac{\alpha}{4\zeta(3)T^2}K_1\(\frac{\muu}{T}\) (\muu^2+2m_e^2)\sqrt{\muu^2-4m_e^2}\ . 
\label{cAverage}
\ee
Using this expression the the final abundance can be approximated by 
\begin{equation}
Y_2 \simeq  1.2\times 10^{17}\mixo^2\, \(\frac{{\rm GeV}}{\muu}\) \left[ \frac{1}{\sqrt{g_{{\rm eff}}} h_{\rm{eff}}} 
\frac{\dd \ln s}{\dd \ln T^3}
\(1+\frac{2m_e^2}{\muu^2}\)\sqrt{1-\frac{4m_e^2}{\muu^2}}\right]_{T=T_d} \ . 
\label{HeavyY}
\end{equation}
Here we have neglected the temperature dependence of the number of radiation degrees of freedom $g_{\rm eff},h_{\rm eff}$ (see appendix \ref{A:dof}) and of the electron mass $m_e$. This is because the integral is dominated by a narrow interval around $T_d \sim \muu/3$ where the former cannot change dramatically (thus we can evaluate them at $T_d$) and the role of $m_e$ in \eqref{cAverage} is minimal. 
To understand this last argument note that, due to the thermal component of the electron mass, there is a maximum temperature $T_c$ that allows coalescence. Writting $m^2_e (T)=m_{e,0}^2+ \pi\alpha T^2 l $ with $l\sim {\cal O}(1)$ we find
\be
T_c = \sqrt{\frac{\muu^2-4m_{e,0}^2}{4\pi \alpha l}} 
\ee
which approaches the limit $T_c \simeq 3.3 \muu$ when $\muu\gg m_{e,0}$. Above this temperature no coalescence is possible (this feature is visible in Fig.~\ref{dY2}).  Only in the limit $T \to T_c$ is there a significant phase space suppression making the square root in \eqref{cAverage} small. Since $T_c$ is much higher than $T_d$, the temperature that dominates production,  and since the thermal electron mass decreases as $T^2$, we do not expect any significant deviation from the RHS of \eqref{HeavyY} (unless, of course, $\muu\sim 2m_{e,0}$ itself).

As we mentioned before, resonant production happens at $T_r\gtrsim \muu/\sqrt{\alpha}$ which is always higher than $T_c$. Therefore coalescence is absent during the resonance. Since at the resonance $\muu=m_\G$ this is of course just another way of saying that the photon thermal mass never exceeds twice the temperature-dependent electron mass in a thermal plasma, and photon decay is kinematically forbidden \cite{Braaten:1990de}.


\subsection{The resonant region $(\muu < 2 m_e)$}
\label{s:resonance}

In the small mass regime with  $\muu < 2 m_e$ electron positron coalescence is kinematically forbidden, and thus 
the biggest contribution to the incoherent production abruptly disappears. For such small masses the enhancement factor in \eqref{ratio}  $\muu^2/\omega \abs  \gg 1$ implying that the resonance
is very peaked, and dominates with respect to the incoherent production.
In this section we develop a simple formula for the resonant production, which is also valid (although not relevant) for higher HP masses.

\begin{figure}[tbp]
\begin{center}
\includegraphics[width=7cm]{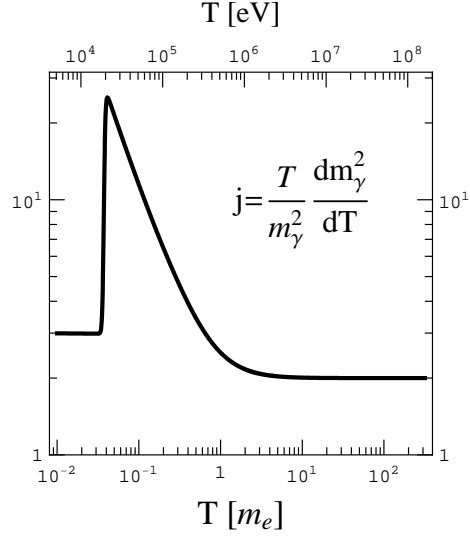}
\caption{Plotted is the function $j = \frac{T}{m_\gamma^2} \frac{\dd m_\gamma^2}{\dd T}$ as a function of the temperature in units of electron mass. The resonant yield $Y_2$ is inversely proportional to $j$, see \eqref{Y2_inter}. }
\label{Fig:j}
\end{center}
\end{figure}

As the resonance time is short, we can neglect the change of plasma
d.o.f. in the Boltzmann equation \eqref{beq}.  If the temperature integral is dominated by the resonance it is easier
to do it before the thermal average of the cross section. Introducing
the notation $\sigma=\chi^2 \hat \sigma$ with
the mixing parameter given in \eqref{mix}, the solution of \eqref{beq} is
\begin{equation}
Y_2 = \sum_{a,b}\int \dd n_a \dd n_b \frac{\hat \sigma_{ab} \MV }{H T s} \frac{\dd \ln s}{\dd \ln T^3}
\frac{\mixo^2\muu^4}{(m_\G^2-\muu^2)^2+(\omega\abs)^2}\dd T
\end{equation}
Next we expand the photon mass around the resonance temperature as
$m_\G^2(T)=\muu^2+m_\G^{2\prime}(T-T_r)+...$; The derivative $m_\G^{2\prime}$ can be expressed as $j(T)\times m_\gamma^2/T$. The function $j(T)$ is plotted in Fig.~\ref{Fig:j}. Limiting cases are $j(T)=3$ for non-relativistic electrons and $2$ in the relativistic case, but notice also the spike at intermediate values due to the steep decline in electron density which suppresses the resonant production. Evaluating all other quantities at their resonance value the integration becomes
\begin{eqnarray}
Y_2 &\approx& \frac{\mixo^2\muu^4}{H T s} \left. \frac{\dd \ln s}{\dd \ln T^3} \int \dd n_{a}  dn_{b} \hat \sigma \MV
\right|_{T=T_r} 
\int_0^\infty\frac{d T}{(m_\G^{2\prime})^2(T-T_r)^2+(\omega\abs)^2}=
\nonumber
\\
&=& \mixo^2 \frac{\pi \zeta(2)}{\zeta(3) }\frac{\muu^2  Y_1}{H j(T) T} \left. \frac{\dd \ln s}{\dd \ln T^3} \right|_{T=T_r}  
\label{Y2_inter}
\end{eqnarray}
To get the second line we used that he temperature integral just gives $\simeq \pi /(\omega {\abs} m_\G^{2\prime})$. This quantity only depends on the outgoing hidden photon energy $\omega$, which allows to reorganize the integrals over particle three-momenta 
\be
\left. \sum_{a,b}  \int d n_a dn_b \frac{1}{\omega{\abs}} \hat \sigma \MV \right|_{\muu=m_\gamma} \equiv \int\frac{g_{{\gamma_{2}}}\dd^3 p_{\gamma_2}}{(2 \pi)^3} \frac{1}{\omega{\abs}} \Gamma^{\rm P}=\int \frac{g_{{\gamma_2}}\dd^3 p_{\gamma_2}}{(2 \pi)^3} \frac{1}{\omega}\frac{1}{e^{\omega/T}-1} =\frac{\zeta(2) }{\zeta(3)}Ê \frac{n_\gamma}{T}
\ee
since $\Gamma^{\rm P}(e^{\omega/T}-1)={\abs}$, and we have used $m_\G=\muu\ll T$ and\footnote{Recall that we do not consider longitudinal modes, which have other dispersion relation.} $g_{\gamma_2}=g_{\gamma_1}=2$ in the thermal integral. Plugging it back in then gives the 2nd expression in \eqref{Y2_inter}.

The cross section (the production rate) has
dropped out of the number density of HPs produced.  This is a curious
result, as the final abundance is independent of the details of the
processes involved, it only depends on the shape of the resonance.

In the limit that the electrons are relativistic, i.e. for temperatures $T\gtrsim m_e$, we can obtain a simple formula through $m_\G^2=\muu^2=2 \alpha \pi T^2/3$,  
\be
Y_2 \approx {1.3} \times 10^{17}\mixo^2\(\frac{\rm GeV}{\muu}\)\mixo^2 \left. \frac{1}{\sqrt{g_{\rm eff}}h_{\rm eff}}\frac{\dd \ln s}{\dd \ln T^3} \right|_{T=T_r}  \ , 
\label{LightY}
\ee
where we set $j(T) = 2$. Up to kinematical factors this expression is the same as the coalescence production yield~\eqref{HeavyY}. Note however that the degrees of freedom are evaluated here at the resonant temperature and in \eqref{HeavyY} at $T_d\simeq 0.3 \muu$. Since the resonance happens at much higher temperatures ($T_r> 8 \muu$) where $g_{\rm eff},h_{\rm eff}$ are larger, its contribution is suppressed with respect to the coalescence yield.

The resulting relic energy density in hidden photons today $\Omega_2$
can be deduced using the conservation of the entropy per coming volume to get 
\begin{eqnarray}
\Omega_2 h^2 = 2.82\times 10^8\ \frac{\mu}{\rm GeV} Y_2\  . 
\label{Omega}
\end{eqnarray}
with $h\sim0.71$ the Hubble constant today in units of $100$ km/s/Mpc. 
Since the values of $Y_2$ \eqref{HeavyY},\eqref{LightY} are inversely proportional to the HP mass (up to a small dependence of the $g_{\rm eff},h_{\rm eff}$ parameters on the production temperature), the relic densities we obtain are practically independent of $\muu$. For our numerical calculations shown in Fig.~\ref{bounds} we used the full expression \eqref{Y2_inter}, rather than some limiting approximation such as \eqref{LightY}.


\section{Bounds}\label{s:bounds}

In this section we discuss the cosmological and astrophysical bounds
on hidden photons with mass and mixing parameter $(\muu,\chi)$.  We do
not make any additional assumptions about the hidden sector, such as
the possible existence of millicharged particles or a hidden sector Higgs.  

Note that in our computation of the relic abundances we have neglected possible backreactions that convert hidden photons into standard model particles. This remains true until $n_2\simeq n_1$, i.e. until $2 \pi^4 h_{\rm eff} Y_2 /(45 \zeta[3])$ is ${\cal O}(1)$. Up to a small mass dependence, this happens for $\mixo\lesssim 10^{-9}$ (see the thin dashed line in Fig.~\ref{bounds}). Of course our computations are not precise outside this regime, but we can qualitatively interpret that for such large kinetic mixing the studied interactions keep hidden photons in thermal equilibrium with the standard bath, at least for a short period of time.

\subsection{Decay rates}

The phenomenology is very different for heavy hidden photons ($\muu > \MeV$) and lighter ones. This has shown up already in the production rate and also affects the decay rate. The mass eigenstate $\gamma_2$ has a
small admixture of the active photon state, and thus can decay.  For
HP masses smaller than two times the electron mass, $\muu < 2 m_e$,
the only decay channel is into three photons through an electron loop\footnote{The $\pi^0$ anomaly also provides a 3 photon decay channel but the amplitude is suppressed by $\Lambda_{\rm QCD}^8$ instead of $m_e^8$ and therefore is subdominant. There is also a decay into two neutrinos, since in general the hidden U(1) should mix with the hypercharge U(1). However, in this case the mixing gets an additional suppression factor $\muu^2/(\muu^2-M_Z^2)^2$ which for the masses considered is huge.}. For larger masses the
dominant decay channel is into an electron-positron pair. The relevant\footnote{Decays are only kinematically allowed for
temperatures below the resonance $(T \lesssim T_r)$ such that the photon and electron thermal masses are smaller than $\muu$.
Since as we will see hidden photon production for $T > T_r$ is
negligible, this threshold is in practice irrelevant.} lifetimes $\tau$ are given by
\bea
\tau^{-1}_{\gamma_2 \too\gamma\gamma\gamma}  &=& 
\frac{17\alpha^4 \chi^2}{11664000 \pi^3}\frac{\muu^9}{m_e^8},
\hspace{4.5cm} (\muu\ll m_e) 
\\
\tau^{-1}_{\gamma_2 \to e^+  e^-} & =& \frac{\alpha\mix^2 \muu}{2}
\sqrt{1-\left(\frac{2 m_e}{\muu}\right)^2}
\left(1+\frac{2 m^2_e}{\muu^2}\right),
\hspace{1.4cm} (\muu > 2m_e) \label{eedecay}
\eea	
with $\alpha \approx 1/137$ the fine-structure constant.  As the HP
photon mass increases, new decay channels into pairs of higher mass charged particles open up. Decays into heavier leptons are analogous to \eqref{eedecay} and the contribution of decays into hadrons are given by $\tau^{-1}_{\rm had}=\tau^{-1}_{\rm e^+e^-}\times R(\muu)$ with $R(s)$ the experimentally measured ratio $\sigma(e^+e^-\to{\rm hadrons})/\sigma(e^+e^-\to{\rm \mu^+\mu^-})$ as a function of the invariant center of energy mass~\cite{Amsler:2008zz}. Unfortunately, as we will see, such heavy HPs cannot be firmly constrained.

If the hidden photon decays before the onset of big bang nucleosynthesis (BBN), about $\sim 0.1\,{\rm s}$ after the big bang, they leave no trace in cosmology.  For lifetimes between BBN and today ($t_0\simeq 4.3\times 10^{17}$ s) the decay products can
affect BBN and/or the CMB blackbody spectrum.
For lifetimes longer than the age of the universe the relic HPs contribute to dark matter so their energy density cannot exceed the fiducial value $\Omega_{\rm DM} h^2\simeq 0.1$. In addition, photons produced in HP decays cannot exceed the measured cosmological photon backgrounds. 

\begin{figure}[htbp]
\begin{center}
\includegraphics[width=13cm]{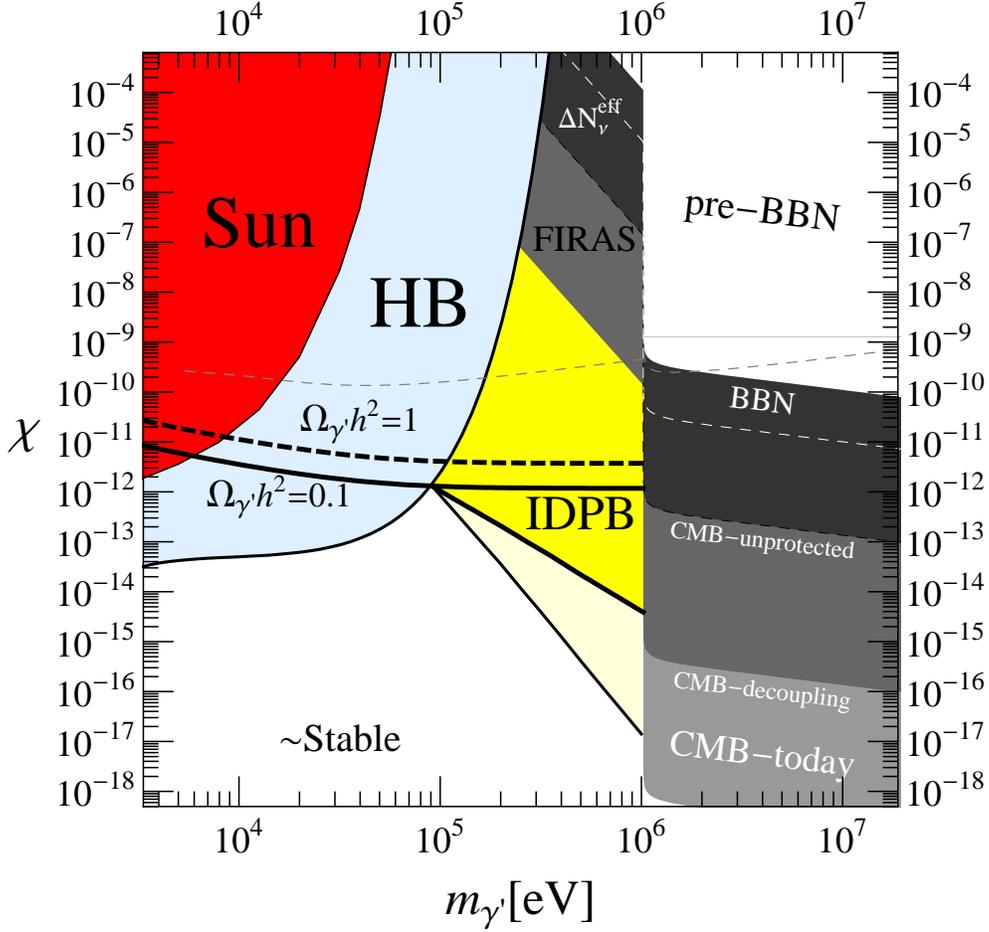}
\caption{Bounds on hidden photons in the mass-mixing plane. 
HPs that reproduce the right amount of DM lie on the line $\Omega_2 h^2=0.1$ and the region above is excluded by overproduction. Above the thin dashed line HPs will interact strongly enough to reach thermal equilibrium with the standard bath.
The regions labeled Sun and HB are excluded by an excessive HP luminosity in the sun respectively in horizontal branch stars in globular clusters.
In the region labeled IDPB the HP decay products exceed the intergalactic diffuse photon background. This bound assumes that the HP relic density is created through the kinetic mixing as discussed in this paper. If one assumes other production mechanisms that lead to $\Omega_2 h^2=0.1$ independently of $\mix$ the bound extends all down to the light yellow region.
We find no bounds above $\muu>2 m_e\simeq 1$ MeV.
Also shown are regions where the HP decay could influence different cosmological epochs: pre-BBN ($\tau<1$ sec), 
BBN ($1$ sec$<\tau<3$ min) and post BBN ($3$ min$<\tau<10^6$ sec), CMB-unprotected ($10^6$ sec$<\tau<10^{12}$ sec), 
CMB decoupled until now ($10^{12}$ sec$<\tau< 4.3\times 10^{17}$ sec ). See the text for details.}
\label{bounds}
\end{center}
\end{figure}

In Fig.~\ref{bounds} we plotted in $(\muu,\chi)$-parameter space the regions for
which the HP decays before BBN (white upper right part),  during
BBN (dark gray band), during the time the CMB is unprotected to distortions (medium gray band)
and after decoupling (light gray band). The region where the HP has a lifetime longer than
the age of the universe, and thus can (if abundantly produced) be 
$\sim$stable dark matter, corresponds to the white lower part of the plot. The
decrease in decay time as the HP mass passes the pair production
threshold is huge. Consequently the most interesting region, where
the HPs can affect cosmology and/or astrophysics, is the low mass
region $\muu < 2m_e$; 
Let us from now on call HPs in this region ``light" HPs. 
They are the main focus of this work as opposed to to ``heavy" HPs which essentially evade all constraints.

\subsection{Cosmological Bounds}

The relic density cannot exceed the measured dark matter fraction in
the universe as measured by WMAP \cite{Dunkley:2008ie,Komatsu:2008hk} which constrains $\Omega_{\rm CDM}h^2 \lesssim 0.1$. 
Hidden photons that saturate this bound and are stable lie in the black solid line in Fig.~\ref{bounds}.
We see that only light hidden photons can satisfy this criterion, and therefore the matter fraction plotted only includes the contribution from the resonance \eqref{LightY}, which as discussed before
dominates the light HP production. 
The space above this line can be excluded because it leads to too much DM abundance. Strictly speaking this bound only applies to hidden photons with the life-time longer than the age of the universe.  However, HPs decaying after CMB decoupling are also excluded, as they leave their trace in the CMB anisotropies. The remaining parameter space above the CMB-decoupling line can be easily excluded as well, but this will be better argued after considering possible constraints on heavy HPs.

Heavy hidden photons decay into electron/positron pairs much before CMB decoupling. One might wonder if the decay products can alter the successful picture we have of the post BBN cosmology.  Let us address this question for HPs of increasing lifetime.

Heavy hidden photons in the right upper white corner of Fig.~\ref{bounds} have lifetimes smaller than $\sim 1.5$ seconds, which correspond in standard cosmology to the moment when the weak charged-current reactions that keep protons and neutrons in thermal equilibrium freeze out (at $T\sim 0.7$ MeV). As a consequence of this freeze out the neutron density is fixed (except for the very slow neutron decay) instead of decreasing exponentially as it would have done if still in thermal equilibrium with protons. Later, nearly all neutrons will convert into $^4$He nuclei, which provides an indirect test of this epoch. This is the earliest time in the history of the universe that we can test  so far.
Therefore, heavy HPs that decay well before do not leave a trace in cosmology.  As we have seen, production of heavy HPs remains active up to $T_d\simeq \muu/3$. From this it follows that  HPs with masses below $\sim 2.1$ MeV are in thermal equilibrium and thus are present as active degrees of freedom during the p-n decoupling (parameter space roughly below the thin dashed line in Fig.~\eqref{bounds}) and they could in principle affect it.

For longer lifetimes the HPs decay during or after BBN acting as extra radiation during the time of BBN affecting the BBN yields as well. Moreover, their decay products can spoil the successful agreement between observations of primordial abundances and theoretical computations (see~\cite{Iocco:2008va} for a excellent recent review). The energetic primaries $e^\pm,\mu^\pm,\pi^\pm ...$ will initiate electromagnetic or hadronic cascades. The former 
can include photons with energy larger than $\gtrsim 2$ MeV which will then photodissociate the light elements; the latter can have a variety of effects like destroying primordial nuclei, inter-converting protons and neutrons or altering the baryon to photon ratio. 

\begin{figure}[tbp]
\begin{center}
\includegraphics[width=12cm]{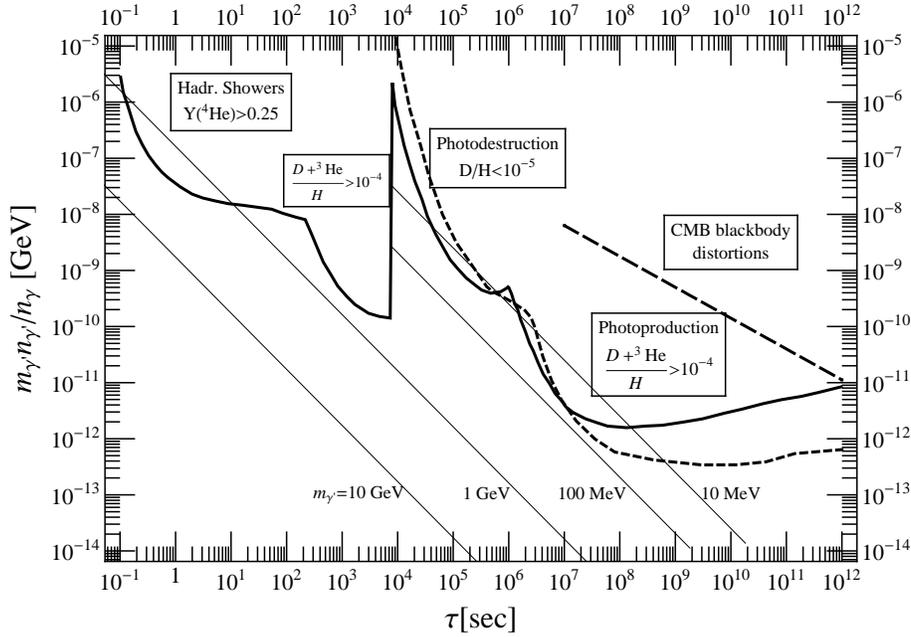}
\caption{Cosmological constraints on a massive particle of mass $\muu$ decaying into SM particles with a lifetime $\tau$ and a relic number density $n_{\gamma'}$. Above the thick solid black line the decay products spoil successful nucleosynthesis~\cite{Ellis:1990nb} producing hadronic showers if $\muu\gtrsim 1$ GeV (below $\tau= 10^4$ s) and electromagnetic cascades if $\muu\gtrsim 5-30$ MeV (above $\tau=10^{4}$~s). The thick short-dashed line comes from a more recent calculation~\cite{Kawasaki:2000qr}. Distortions of the CMB spectrum would be noticeable above the long-dashed line~\cite{Ellis:1990nb}. 
Our predictions for hidden photons for different masses $\muu> 2m_e$ are shown as thin lines.}
\label{BBNsummary}
\end{center}
\end{figure}

Usually one derives bounds on the amount of energy released per photon (here $\muu Y_2  s/n_{\gamma_1}$) 
since the electromagnetic/hadronic cascades form a universal decay spectrum \cite{Ellis:1990nb,Holtmann:1998gd,Kawasaki:2000qr}  independent of the energy of the primaries. 
This turns out to be an excellent approach for very energetic primary particles, but fails for the lowest masses we want to consider of order MeV. For decay photons and electrons below the threshold for this universal cascade, but above the threshold for photodissociation, the resulting spectrum will be different. 
The authors of \cite{Kawasaki:1994sc} considered specific bounds for a late injection of $10$ MeV photons from a relic decaying particle and found them to be less stringent than for universally cascading photons/electrons. They quote the bound $E_\gamma n <10^{-9} \GeV$   for decay photons with energy $E_\gamma< 20$ MeV from a relic particle whose lifetime is $> 10^6$ s and is present with a relic density $n$ relative to the thermal photons.
The bounds present in the literature depend on the plasma interactions considered and (slightly) on the experimental values of the abundances taken. In Fig.~\ref{BBNsummary} we have reproduced the results of \cite{Ellis:1990nb,Kawasaki:2000qr} as an illustration and we have plotted the HP energy release per photon $\muu Y_2 s/n_{\gamma_1}$ as a function of the Hidden Photon decay lifetime $\tau$.

The region above the thick line below $\tau=10^4$ s is excluded for relics with masses $> 1$ GeV which can trigger hadronic cascades via their decay products. The line $\muu = 1$ GeV cuts this line for an abundance which corresponds to $\mixo\sim 10^{-11}$, but this mass is on the borderline of the limit of validity of this bounds. A more dedicated analysis is needed to potentially rule out a (small) region around these values. Something very similar happens for lifetimes above $10^4$ s. Here the decay products are limited by their capability of initiating electromagnetic cascades which contain photons of energy above $2.2$ MeV that can destroy deuterium, or above $20$ MeV that can destroy $^4$He giving as a product too much Deuterium and $^3$He. We have also plotted the more recent results of \cite{Kawasaki:2000qr} as a dashed line. The authors of \cite{Ellis:1990nb} claim the bounds to be valid for primary photons/electrons above $5-30$ MeV. Again we have a small region which can be be potentially excluded, but it is once again at the limit of validity of the bound, and a more carefull analysis is needed.  

To summarize, in the high mass region $\muu > 2m_e$ we find no significant bounds from BBN. Basically, if the mixing is large so that appreciable abundances of HPs are produced, the decay width is
also large and decay occurs before it can cause troubles to our standard picture of BBN.

Let us now comment on possible distortions of the CMB blackbody spectrum. A perfect blackbody has fixed energy and number densities of particles which are governed by a single parameter, the temperature. The decay products of hidden photons can change the black body spectrum. The photon interactions with the ambient electrons and nucleons will tend to bring them back to their blackbody values, although at a different temperature. Compton scattering is effective in redistributing energy until the universe is $\sim10^9$ seconds old, but cannot change the photon number. 
The interactions responsible for this are bremsstrahlung $e^-p^+\to e^- p^+\gamma$ and double Compton scattering $e^-\gamma\to e^-\gamma \gamma$, and they become ineffective at around $10^6$ seconds at $\sim$ keV temperatures. 
Therefore, if the HP decays before $10^6$ seconds the blackbody will be eventually reestablished. If it decays between $10^6$ and $10^9$ seconds the photon spectrum will regain a thermal shape, but with a non-zero chemical potential. Finally if the decay is later than $10^9$ seconds there will be a noticeable non-thermal distortion in the CMB called ``Comptonization" or y-distortion \cite{Zeldovich:1969ff}. These last two  possibilities can be strongly constrained by the precise determinations of the blackbody spectrum of the CMB by the FIRAS spectrometer on board of the COBE satellite \cite{Fixsen:1996nj}. The bounds on the energy injected were derived in \cite{Ellis:1990nb} and are also shown in Fig.~\ref{BBNsummary} as a long dashed line. 
We again see that heavy hidden photons cannot produce noticeable distortions. 

Let us now examine BBN and CMB bounds for light HPs ($\muu<2 m_e$). They have masses below the electron mass, so they cannot produce photons with energy greater than 2 MeV which could photodissociate deuterium. In Fig.~\ref{bounds} we already see that light HPs that decay before the CMB decoupling lie above the thermalization line, so during their resonant production at $T_r> 8 \muu$ they reach thermal equilibrium with photons. This fixes their abundance rather than \eqref{LightY}. Later, when they decay, their energy density is \emph{at least} as large as that of the photons. Using $\muu \sim m_e$ and  $n_{\gamma_2}\sim n_{\gamma_1}$ and examining the CMB bounds of Fig.~\ref{BBNsummary} it becomes clear than the region labeled FIRAS in Fig.~\ref{bounds} is completely excluded. 
Light hidden photons with larger kinetic mixing will decay when photon-number interactions are active and therefore will not distort the spectrum. However, if they decay after BBN (above the white dashed line) they will certainly distort standard BBN acting as invisible degrees of freedom. Since they are in thermal contact until $T\sim 8\muu \sim 5$ MeV and then suddenly decouple they miss the injection of entropy of electron positron annihilation, and they contribute to the energy density during BBN as 8/7 that of a neutrino species. At least, this would be the case if they were massless. Kolb and Scherrer showed \cite{Kolb:1981cx} that a neutrino with mass in the range $0.1-10$ MeV affects BBN more strongly than a massless neutrino, and the same conclusion remains valid for our light HPs. Similar conclusions can probably be drawn even if they decay before BBN although a detailed study would be required for this bounds to be robust. Finally if $\mixo\sim 10^{-3}\sim 10^{-4}$ they would decay immediately after their resonant production, before p-n decoupling leaving no trace in BBN.

\subsection{Gamma Ray Bounds}

The decay of light dark matter hidden photons 
$\gamma_2 \to 3\gamma$ would contribute
to the observed gamma-ray background. Bounds on decaying
dark matter have been extensively discussed in the 
literature~\cite{DeRujula:1980qd,Berezhiani:1987gf,Doroshkevich:1989bf,Berezhiani:1990jn,Khlopov:1997bk,Bertone:2007aw,Zhang:2007zzh}.  
Most of these works concentrate on 2-body decay.  Line searches in the spectrum are
done to bound the decay rate and mass of the decaying DM
particle.  Hidden photons decay into three photons, and the resulting
spectrum has no sharp line feature.  Instead of applying the line
searches to our 3-body decay spectrum, we demand that the gamma
ray flux from decaying HPs does not exceed the total observed
gamma-ray flux\footnote{The bound obtained is a factor $10^{-2}$ weaker than that
obtained for line searches for 2-body decay.  
A similar suppression factor is obtained when we try to adopt the line-search result to 3-body
decay, as the maximum hits per bin is reduced by a factor $10^{-2}$ with respect to the 2-body decay.}. 
Model-independent bounds were calculated in~\cite{Yuksel:2007dr} imposing that the photon flux from the DM decays does not exceed the measured intergalactic diffuse photon background (IDPB):
\be
\frac{\muu \, \tau}{\rm GeV s} \lesssim 10^{27}\(\frac{\omega}{\rm GeV}\)^{1.3}  \( \frac{\Omega_2 h^2}{0.1}\)
\label{YukselBound}
\ee
for a particle of mass $\muu$, lifetime $\tau$ and $\omega$ the energy of the decay photon. The energy spectrum of the decay photons is peaked around $\omega \simeq \muu/3$.  Ref. \cite{Yuksel:2007dr} assumed $\Omega h^2=0.1$, setting the last factor in \eqref{YukselBound} to unity. 
The used energy bin width is large enough so that the bounds for 2-body and 3-body decay are nearly the same. 

In Fig.~\ref{bounds} we plotted the curve in $(\muu,\chi)$-plane
for which the HP decay is 10\% and 100\% of the total observed flux.
The parameter space above this line is excluded.  In our calculation we did not assume that HPs are all of the dark matter setting $\Omega_2 h^2 =0.1$ in the RHS of \eqref{YukselBound}, \emph{but instead calculated the relic abundance as a function of ($\mu,\chi$) from our calculations of section~\ref{s:production}}.

\subsection{Astrophysical Limits}

The production of weakly interacting particles in stellar interiors
can substantially affect stellar evolution \cite{Raffelt:1996wa,Raffelt:2006cw}.
Hidden photons with small $\mix$ will be scarcely produced in the
dense plasmas of the stellar interiors, but they will
leave the star unimpeded contributing directly to the star's overall
luminosity. On the other hand, only photons of the photosphere (and neutrinos) can contribute to the
standard energy loss. Therefore, naively, the hidden photon luminosity
is enhanced at least by a volume/surface factor and a further
$(\rho_{\rm inside}/\rho_{\rm surface})^n(T_{\rm inside}/T_{\rm surface})^m$ (with $\rho$ a typical particle density and $n,m>1$) with respect to the photon luminosity.

The effect of such an energy loss in the case of the Sun was studied
in \cite{Redondo:2008aa}. In a main sequence star like our Sun the
existence of an exotic luminosity simply means that the fusion of
hydrogen into helium is proceeding faster than required to account
only for the observed photon luminosity. It is well known that the Sun
has been shining for more than 4 billion years, and this constraints
the rate of helium production (since there is still some H in the
Sun). This translates into a bound ${\cal L}_\GP < {\cal L}_\gamma$
for the hidden photon luminosity. Integrating over a solar model
\cite{Bahcall:2004pz} gives the bound labeled Sun in
Fig.~\ref{bounds}. An improvement of roughly one order of magnitude in
the bound can be achieved by matching the photon luminosity with the
neutrino flux, also proportional to the rate of hydrogen fusion, as
done for the axion case in \cite{Schlattl:1998fz}. However we already
see that the bound degrades very fast for $\muu \gg T_\odot\sim 1.3$
keV, leaving space for hidden photon dark matter. To close the window
we need to consider hotter stellar interiors.

We shall focus on horizontal branch (HB) stars in globular clusters which
provide the strongest limits for energy loss for intermediate
temperatures. The core of a HB star burns helium into carbon and oxygen at a quite constant temperature of $8.6$ keV and density $\sim 10^4$ gr
cm$^{-3}$. With these characteristics a HB core is still a
classical plasma and the Compton process provides the most important HP production mechanism. 
At such small temperatures the thermally averaged production cross section 
can be computed precisely taking into account the correct statistics for the photon. 
The invariant mass is practically independent of the electron momentum 
$s\simeq m_e^2+2m_e \omega$ so $\int \dd n_e$ factorizes out. Also the relative velocity is simply $\MV \sim 1$. We find for the energy loss per unit mass
\be
\frac{\dd E}{\dd m\dd t} = \frac{1}{2 m_u}\frac{1}{\pi^2}\int_\muu^\infty\frac{\omega^3 \dd\omega}{e^{\frac{\omega}{T}}-1}\sigma_{\gamma_2 e}(s(\omega)) \ , 
\ee
which has to be smaller than $10$ erg g$^{-1}$ s$^{-1}$
\cite{Raffelt:1996wa}. Note that in a HB core the plasma frequency at the core is $\omega \sim 2$ keV so we can take
$\mix_{\rm eff}\simeq \mix$ for masses $\muu\gg 2$ keV. The resulting bound is plotted in Fig.~\ref{bounds}.

Usually hotter stars are also more dense and, having larger plasma
frequencies, it is possible to have a resonant production for $\muu> 2$ keV. This can be the case for red giants before He ignition ($T \sim 8.6$ keV, $\omega \sim 18$ keV) and white dwarfs ($T \sim 1$ keV, $\omega \sim 23$ keV).
All these stars have cores supported by electron (or neutron) degeneracy pressure. 
Photon interaction rates in these environments are suppressed by Pauli blocking factors of fermions, making the HP energy loss calculation more involved. 
Moreover, the HB bound constraints HPs as a sizable fraction of dark matter for $\muu\lesssim 100$ keV. 
Only in supernovae cores (up to $T \sim 10$ MeV, $\omega \sim 1$ MeV) can the density  be so high to get resonant production of such massive HPs. Unfortunately, in this case we find that the emission of HPs cannot compete with neutrinos, and the bounds are not competitive with the IDPB bound described above \cite{Pospelov:2008jk}.

\section{Non-renormalizable operators}
\label{s:non-ren}

In our treatment so far we assumed that kinetic mixing is the only relevant coupling between the standard model and the hidden sector. 
Kinetic mixing can be generated at arbitrary high energy scale by ``messenger fields'' which are charged under both the SM U(1) and hidden sector U(1). However, these same messenger fields generate at one-loop order also a non-renormalizable term in the Lagrangian. It is not impossible that these additional interactions dominate the production of hidden photons. If so, this opens up parameter space and allows for the possibility of HPs with mixing parameter smaller than $\chi \sim 10^{-11} - 10^{-12}$ as warm dark matter (hidden photons with larger mixing are still excluded as they overclose the Universe). Let's discuss this scenario in a bit more detail.

For definiteness, consider two messenger fields with masses $M,M'$ and charges $(1,1)$ and $(1,-1)$ under weak hypercharge and the hidden U(1) respectively. The kinetic mixing generated at one loop is then~\cite{Holdom:1985ag}
\be
\chi = \frac{g_Y g'}{16\pi^2} \log\(\frac{M'^2}{M^2} \)
\approx \frac{g_Y g'}{16\pi^2} \( \frac{\delta M^2}{M^2} \)
\sim 10^{-4} \( \frac{\delta M^2}{M^2} \).
\label{chi_UV}
\ee
In the second step we assumed degenerate masses $\delta M^2 = M'^2 - M^2 \ll M^2$. In the last step we took the hidden sector gauge coupling $g'$ to be of the same order as the hypercharge coupling $g_Y$, with $1/60 = \alpha(M_Z) < \alpha < \alpha({\rm GUT}) < 1/25$.
To get small enough kinetic mixing $\chi \lesssim 10^{-11}$ requires very degenerate messenger fields $\delta M/M \lesssim 10^{-8}$, very small $g'$, or a combination of both. 

The same messenger fields also give rise to non-renormalizable operators in the Lagrangian, for example (up to order one factors) 
\be
\mcL \supset \frac{g_Y^2 g'^2}{(4\pi)^2 M^4 } (F_{\mu \nu} F^{\mu \nu})
(B_{\rho \sigma} B^{\rho \sigma}).
\label{non_ren}
\ee
A significant amount of hidden photons can be produced through the above interaction provided the reheat temperature of the universe is high enough. Since the HP production rate $\sim \alpha \alpha' (T/M)^8 T$ depends on $T$ to the ninth power,  for temperatures $T \ll M$ these processes are negligible, and kinetic mixing dominates the production of HPs.
The non-renormalizable interactions are in thermal equilibrium in the early universe if the interaction rate exceeds the Hubble rate $H \sim T^2/\mpl$. This happens for
\be
\( \frac{T}{M} \)^7 \gtrsim
\frac{1}{\alpha^2 \alpha'^2}\( \frac{M}{\mpl}\)
\ee
that is for reheat temperatures $T_{\rm RH} \gtrsim 0.1 M$.  If equilibrium is established the number density of hidden and SM photons is of the same order of magnitude at the time the reactions \eqref{non_ren} freeze out. Rewriting to entropy density, and using \eqref{Omega} this translates into a HP relic density of 
\be
(\Omega_2)_{\rm eq} \simeq 8\times 10^2 \(\frac{\mu}{\MeV}\)
\( \frac{200}{h_{\rm eff}}\)
\ee
with $h_{\rm eff} \gtrsim 200$ the effective entropy degrees of freedom (see Appendix~\ref{A:dof} ) at freeze-out. The dark matter density $\Omega_2 h^2= 0.1$ is obtained for masses in the  $\mu \sim $ keV range.

For larger HP masses, the right relic density can only be obtained if production occurs out of equilibrium. Production is most effective at the largest temperature which is the reheat temperature. The estimated relic density is \cite{Chen:2008yi}
\be
(\Omega_2 h^2)_{\rm non \; eq} \sim 
6 \times 10^{-3} {g'}^4 \( \frac{\mu}{\keV} \)
\( \frac{T_{\rm RH}}{M}\)^7 \( \frac{\mpl}{M} \)
\ee
which can easily be of order $0.1$ for a wide range of masses, depending on the parameters $g',T_{\rm RH},M$. Without an UV completion of the theory no definite predictions can be made. The result is independent of the mixing parameter. From \eqref{chi_UV} it can be seen that the mixing parameter depends on the unknown mass degeneracy $\delta M/M$ of the messenger fields.

The constraints on this scenario are the same as when kinetic mixing is the only production source. As mentioned before, the mixing parameter should be smaller than $\chi \lesssim 10^{-11}-10^{-12}$ to avoid overclosure. The resulting HPs have a lifetime larger than the universe, and can be the dark matter. The stellar constraints from the sun and horizontal branch stars are unchanged, but the gamma ray bounds are slightly different than before. In Fig.~\ref{bounds} we used the HP parameters $\{\mu,\chi\}$ to calculate the relic density produced via kinetic mixing, which in turn was used to determine the gamma ray bound for the parameters at hand. In the present scenario the relic density gets an additional and unknown contribution from early production via the non-renormalizable interactions.  Therefore in Fig.~\ref{bounds} we plotted the gamma ray bounds taking a fixed value of the relic density $\Omega_2 h^2= 0.1$, as well as the bounds when $\Omega_2 h^2$ is computed from production via kinetic mixing and thus depends on $\{\chi,\mu\}$. Parameters above the gamma-ray bound for which $\Omega_2 h^2$ is dominated by kinetic mixing are still excluded. Parameters in between the two lines are excluded as the dark matter candidate, but for smaller relic densities the gamma ray bounds are still evaded. Finally, the parameters below both bounds are unconstrained by the $\gamma$-background.

The discussion so far concerned hidden photons with masses $\mu < 2m_e$ which could be the dark matter. Of course, the production of heavier HP can also be enhanced by the non-renormalizable interactions. The resulting constraints from the CMB and BBN are very model dependent, as they depend on the unknown parameters $M,\delta M,T_{\rm RH}$. The constraints for a fixed number density can be found in the literature \cite{Ellis:1990nb,Holtmann:1998gd,Kawasaki:2000qr,Iocco:2008va}.

\section{Conclusions}

In this paper we studied the production and cosmology for hidden photons with a mass in the keV-MeV range. The main motivation for the undertaking of this project is that such hidden photons can be the lukewarm dark matter of the universe.

In our study we focused on the case that the main interaction between the hidden sector and the SM is via kinetic mixing with the photons. In particular the relic abundance is computed under the assumption that kinetic mixing is the dominant production mechanism. For a proper calculation all relevant plasma effects to the photon's self-energy have to be taken into account. The net result is that for sub-MeV hidden photons the production is dominated by the resonance which occurs at temperatures such that the plasma mass of the photon equals the hidden photon mass. The final abundance is almost independent of the HP mass. 
The dark matter abundance can be obtained for mixing angles of order $\chi \sim 10^{-11}-10^{-12}$. For heavier, super-MeV hidden photons, the main production channel is via electron-positron coalescence.

Hidden photons are an example of decaying dark matter. Light HPs can decay into 3 photons via the kinetic mixing interaction; heavier HPs with a mass $\mu > 2m_e$ can also decay into a electron-positron pair with a relatively large decay rate. For the HPs to be the dark matter its lifetime should be of the order of the age of the universe or larger. This excludes the heavier HPs with super-MeV masses. What is even more, we showed that such heavy HPs leave practically no imprint on cosmology even if they decay after the time of nucleosynthesis and/or decoupling. This is different for sub-Mev hidden photons. Indeed, light HPs with mixing parameter $\chi \sim 10^{-11}-10^{-12}$ are stable enough to be dark matter. But they do decay, and the decay photons contribute to the diffuse gamma-ray background. As this background is measured in the relevant energy range, this puts bounds on the HP relic density, and thus on the mass and mixing parameter. We find that HP with masses $\mu \gtrsim 100$ keV are excluded as dark matter.

Apart from the bounds on the relic density, the lifetime of the HPs, and the diffuse gamma ray background there are in addition bounds from stellar evolution. If light enough, HPs are copiously produced inside stars. As their interaction rate is small, they subsequently leave the star (almost) unimpeded. Thus HPs present an effective cooling mechanism for stars, which is bounded by observations. For our study in particular the bounds from the Sun and horizontal branch stars in globular clusters are important; they exclude all parameter space for masses $\muu \lesssim 100$ keV. 

All bounds put together exclude HPs as main source of dark matter. There is a loophole in this argument though. We assumed that production is dominated by kinetic mixing. But if the reheat temperature of the universe is high enough non-renormalizable operators, which are always present, can play a role too. In particular, if HPs are pre-dominantly produced via these non-renormalizable interactions, parameter space opens up, and HP can still be the dark matter in the universe. Even if this is not the case, HPs can provide a fraction of the dark matter, which can be testable in the future through its 3-photon decay.

\section*{Acknowledgments}
JR would like to thank A.~Ringwald and C.~Weniger for fruitful conversations, and Nikhef for its hospitality. MP is supported by a VIDI grant from the Dutch Organisation for Scientific Research (NWO). 

\appendix

\section{Photon self energy in the primordial plasma} \label{A:wp}

The photon self-energy in the early universe plasma receives its lowest order contribution in $\alpha$ from Compton scattering off the thermal electrons \cite{Altherr:1992mf,Braaten:1993jw}. It turns out to be different for transverse ($\pi_{\rm T}$) and longitudinal excitations ($\pi_{\rm L}$). In this paper we only focus on the first case.
The computation of the real part is nicely reviewed in~\cite{Raffelt:1996wa}. 
One finds that it is closely related to the plasma frequency $\OP$, 
\bea
\label{ReT}
{\rm Re}\ \pi_{\rm T}=m_\gamma^2 &\simeq & \OP^2\(1+\frac{1}{2}G\(v_*^2k^2/\omega^2\)\)\\
\OP^2 & = & 
\frac{4 \alpha}{\pi}\int_{m_e}^{\infty}\( f_{e^+}+f_{e^-} \) \(v-\frac{1}{3}v^3\)e d e
\eea
with $k,\omega$ the photon momentum and energy, $e$ the electron energy, $f_{e^\pm}$ the Fermi-Dirac distributions of electrons and positrons and the $e^\pm$ velocity is $v=\sqrt{1-m_e^2/e^2}$. Here $v_*$ is a typical electron velocity given by $v_*=\omega_1/\OP$ and $\omega_1$ and $G(x)$ are defined through
\bea
\omega_1^2=\frac{4 \alpha}{\pi}\int_{m_e}^{\infty}\( f_{e^+}+f_{e^-} \) \(\frac{5}{3}v^3-v^5\)e d e  \\ 
G(x)=\frac{3}{x}\[1-\frac{2x}{3}-\frac{1-x}{2\sqrt{x}}{\rm Log}\(\frac{1+\sqrt{x}}{1-\sqrt{x}}\)\] \ .
\eea
Note that $G(x)$ ranges from $0$ to $1$, so in practice the effective mass is never far from the plasma frequency.
In the cases we are interested in photons have energies much larger than $\OP$ and therefore one can take $k\simeq\omega$ in \eqref{ReT}. The effective mass is then effectively independent of $\omega$.

The imaginary part is given by \eqref{abs} though the difference of the photon production $\Gamma^{\rm P}$ and absorption rates $\Gamma^{\rm A}$. Considering the reactions $\{i\}\to \{f\}+\gamma$ where $\{i\},\{j\}$ are sets of possible initial/final states they are given by
\bea
\Gamma^{\rm P}(\omega,T)=\frac{1}{2\omega}\sum_{\{i\}}\sum_{\{f\}}\int d\Omega_{if}|\tilde {\cal M}(\{i\}\to \{f\}+\gamma )|^2 
\(\prod_{a\in \{i\}} f_a\) \(\prod_{b\in \{f\}} (1\pm f_b)\)\\
\Gamma^{\rm A}(\omega,T)=\frac{1}{2\omega}\sum_{\{f\}}\sum_{\{i\}}\int d\Omega_{if}|\tilde {\cal M}(\gamma+\{f\}\to \{i\} )|^2 
\(\prod_{a\in \{f\}} f_a\) \(\prod_{b\in \{i\}} (1\pm f_b)\)\\\eea
where $\cal \tilde M$ are spin-averaged matrix elements, $f_x$ are either Fermi/Bose distributions, the $\pm$ is to be chosen to account for the stimulated boson emission ($+$) or for the fermion blocked emission ($-$). The phase space integration is given by
\bea
d\Omega_{if}=(2\pi)^4\delta^4(K^\mu+\sum_{a\in \{i\}} p^\mu_a-\sum_{b\in \{f\}} p^\mu_b)\(\prod_{a\in \{i\}} \frac{g_ad^3p_a}{(2\pi)^32E_a}\)\(\prod_{b\in \{f\}} \frac{g_bd^3p_b}{(2\pi)^32E_b}\)
\eea
where $p^\mu_x(p_x)$ and $E_x$ are the 4(3) momenta and energy of the partice species $x$ and $g_x$ the number of internal degrees of freedom.

\section{Degrees of freedom in the early universe}\label{A:dof}

The expansion rate is governed by the Hubble parameter, $H$, which neglecting curvature effects is given by
\be
H^2=\frac{8 \pi G_{_{N}}}{3} \rho
\ee
where $G_{_{N}}$ is Newton's constant and $\rho$ is the universe energy density. Usually one expresses the total energy and entropy densities $\rho,s$ by normalizing them with the contribution of a thermalized relativistic bosonic species getting the so-called effective number of species
\be
g_{\rm eff} (T) \equiv \frac{30}{\pi^2 T^4}\ \rho \BSP h_{\rm eff} (T) \equiv \frac{45}{2\pi^2 T^3} \ s \ .
\ee
The factor 
\be
\frac{\dd {\rm Log}s}{\dd {\rm Log} T^3}=1+\frac{1}{3}\frac{T}{h_{\rm eff}}\frac{\dd h_{\rm eff}}{\dd T}
\ee
is different from $1$ only near the annihilation thresholds $T\sim m_x$ of standard model particles.

\section{Cross sections for hidden photon production}\label{A:cs}

Hidden photons interact though their kinetic mixing with photons with all standard model charged particles, i.e. though the electromagnetic current $j^{\rm em}_{\mu}$ as
\be
j^{\rm em}_{\mu} A^\mu \simeq j^{\rm em}_{\mu}\(A_1^\mu-\mix A_2^\mu\) \ \ .
\ee
Therefore for every SM process in which a photon is emitted, there is also a HPs production channel with a coupling reduced by $\mix(\omega,T)$ given by \eqref{mix}.

The cross section of electron-positron coalescence into a HP is given by
\be
\sigma_{\gamma_2}(s) = 4\pi \alpha^2\mix^2
\sqrt{s-4m_e^4}\(1+\frac{2m_e^2}{\muu^2}\)\delta(s-\muu^2)
\ee
with $\delta(s-M^2)$ a Dirac delta function. Recall $\mix$ is a function of $\omega$ which in this case is $\omega=\sqrt{s}$.

The cross section of $e^+e^-$ annihilation into a (massless) photon and a HP is given by
\be
\sigma_{\gamma_2\gamma_1}(s) =\frac{4\pi \alpha^2\mix^2}{s-\muu^2} 
\(2\frac{s^2+4sm_e^2 -8m_e^4+\muu^2(\muu^2-4m_e^2)}{s(s-4m_e^2)}{\rm Log}\[\frac{\sqrt{s}+\sqrt{s-4m_e^2}}{2m_e}\]
-\frac{s(s+4m_e^2)+M^4}{s\sqrt{s(s-4m_e^2)}}\)
\ee
where $\mix=\mix(\omega,T)$ from \eqref{mix} and $\omega=(s+\muu^2)/(2\sqrt{s})$.
If the HP mass is big enough ($\muu>2 m_e$) then we find t and u-channel divergences at the threshold $s=\muu^2$, reflected in the denominator of the first factor on the right hand side. In a thermal bath, these divergences are cut off by the photon thermal mass. 
Since the main contribution to the non-resonant HP production comes from temperatures smaller than the resonance $\muu \gg m_\G$ 
and the inclusion of $m_\G$ affects the cross section mainly in the divergent electron propagator, which we can correct applying the prescription   
\be
s-\muu^2\to s-\muu^2+2 m_\gamma \muu
\ee
to the first term.

Finally, the cross section of Compton-like production of hidden photons is given by the more complicated formula
\bea
\sigma_{\gamma_2 e}(s) =\frac{2 \pi \alpha^2\mix^2}{(s-m_e^2)^3}\(
\frac{\beta}{2s}\(s^3+15s^2 m_e^2-sm_e^4+m_e^6+
\muu^2\(7s^2+2 s m_e^2-m_e^4\) \) + \right. \\
\left. 
+2\(s^2-6s m_e^2-3m_e^4-2\muu^2(s-m_e^2-\muu^2)\){\rm Log}\[\frac{s(1+\beta)+m_e^2-\muu^2}{2m_e\sqrt{s}}\] \)
\eea
where $\beta$ is an invariant defined by $s^2\beta^2=(s-(m_e+\muu)^2)(s-(m_e-\muu)^2)$, 
which is proportional to the HP momentum. In this case the HP energy is $\omega=(s+\mu^2-m_e^2)/(2\sqrt{s})$.



\providecommand{\href}[2]{#2}\begingroup\raggedright\endgroup

\end{document}